\documentclass{caosp310}

\usepackage{graphicx}
\usepackage{natbib}
\usepackage{amsmath}
\newcommand{\OIII}{[O~{\sc iii}]}
\newcommand{\HeII}{He\,{\sc ii}}
\newcommand{\NII}{[N\,{\sc ii}]}
\newcommand{\SII}{[S\,{\sc ii}]}
\newcommand{\Ha}{H$\alpha$}
\newcommand{\Hb}{H$\beta$}
\newcommand{\kms}{\,\mbox{km}\,\mbox{s}^{-1}}

\articleNo{361}
\pubyear{2025}
\volume{54}
\volnumber{3}
\firstpage{1}
\received{October 20, 2025}
\accepted{October 28, 2025}

\def\BibTeX{{\rm B\kern-.05em{\sc i\kern-.025em b}\kern-.08em
             T\kern-.1667em\lower.7ex\hbox{E}\kern-.125emX}}

\begin{document}

\hauthor{A.\,Arshinova, K.\,Sanderson and A.\,Moiseev}
\title{Green Bean galaxies and the fading echoes of AGN activity}
\author{
        A.\,Arshinova\inst{1}\orcid{0009-0002-9297-8809}
      \and
        K.\,Sanderson\inst{2}\orcid{0000-0002-0507-9307} 
      \and 
        A.\,Moiseev\inst{1,3}   
       }

\institute{
           Special Astrophysical Observatory, Russian Academy of Sciences,
           Nizhny Arkhyz 369167, Russia \\ 
           \email{arina.arshinova@gmail.com}
         \and 
           Department of Astronomy, New Mexico State University,
           Las Cruces, NM 88033, USA 
         \and
           Space Research Institute Russian Academy of Sciences, 84/32 Profsouznaya str., Moscow, 117997, Russian 
          }

\date{March 8, 2003}

\maketitle

\begin{abstract}
Green Bean  is a rare type of galaxy which represents a short-lived phase in the life cycle of active galactic nuclei (AGN), characterised by large-scale, powerful ionised clouds in the circumgalactic medium. Recent studies demonstrate that these extended ionised structures may reflect fading signatures of past AGN activity, often manifested in the form of large-scale ionisation cones. The analysis of their  observational properties provides unique constraints on AGN lifetimes, feedback mechanisms, and transitions between radiative and kinetic modes of activity. In this paper we announce the first results of the project dedicated to the long-slit  spectroscopic and scanning Fabry-Perot interferometric observations of Green Bean galaxies at the Russian 6-m telescope with SCORPIO-2 multi-mode instrument. We describe the data reduction and spectral fitting procedures that allow one to characterise ionisation conditions in extended gaseous regions of the galaxy SDSSJ095100.54+051026.7. 
\end{abstract}
 
\section{Introduction}
\label{intr}

Active Galactic Nuclei (AGN) represent a critical transitional stage in galaxy evolution, where supermassive black holes undergo intense but temporary phases of accretion. While extremely luminous, their true significance lies in this fleeting yet transformative period -- a key link in understanding how galaxies and their central black holes co-evolve. Although estimates of the duration of the active phase vary and are subject to debate, they have been determined in recent years using a variety of `galactic archaeology' methods \citep[][and references therein]{Morganti2017}. In some radio-loud AGNs, extended relic structures can be identified, marking the decline of past activity and characterised by steep radio spectra and specific morphological features \citep{, Morganti2024Galax..12...11M}. These relics typically trace AGN activity cycles on timescales of $\sim10^{6}$--$10^{8}$~yr, reflecting the gradual fading of synchrotron-emitting plasma. Similarly, in the optical domain, extended photoionised nebulae (ionisation cones) are observed, providing clues to the timescales of their activity. Such structures often extend up to 3--50~kpc, corresponding to light-travel timescales of $\sim10^{4}$--$10^{5}$~yr, while the recombination timescales of the ionised gas are much shorter \citep[see Section~6.5][]{Schirmer2016}, thereby probing much more recent phases of AGN activity.

These indicators of fading activity are typically studied separately for radio-loud and radio-quiet AGNs, with links to different accretion rates --- either low or high, approaching the Eddington limit -- or to kinetic and radiative forms of activity, respectively \citep{Mullaney2013, Kukreti2024}. Recent multi-wavelength studies  of nearby AGNs have revealed rare instances of galactic nuclei transitioning between kinetic and radiative regimes, or vice versa \citep{Harvey2023MNRAS.526.4174H,Moiseev2025}. While a comprehensive theoretical explanation for this behaviour is still lacking, some models suggest that the influence of the nucleus on the interstellar medium during the earlier active phase of the galaxy plays a critical role \citep{Morganti2017FrASS...4...42M, Ciotti2017}.

`Green Bean' galaxies (GBGs), thought to be remnants of quasar ionisation, are extended ionised clouds detected in the SDSS sky survey at redshifts up to $z~\sim~0.6$ \citep{Schirmer2013}. While these systems were originally characterised primarily by their extended emission-line regions, recent work has demonstrated that they can also host extended radio structures, including jet-induced lobes on scales of hundreds of kiloparsecs \citep{Sanderson2024}. Moreover, the same study revealed that the radio jets are at least $6$ Myr old, and estimate of the ionisation balance indicates that the central activity has faded over the past $0.15$ Myr. The ability to probe these accretion and radiative fading timescales makes GBGs ideal laboratories for studying changes in AGN activity, with Hanny’s Voorwerp \citep{Lintott2009} representing a well-studied example in the local Universe. Motivated by this, we have initiated observations of the ionised gas in GBGs already known to host extended radio structures, in order to probe whether the ionisation, kinematics, and spatial extent of the ionised gas correlate with the relic radio morphology, and to characterise the current evolutionary state of the AGN. As part of this program, we use long-slit (LS) spectroscopy and scanning Fabry–Perot interferometry (FPI) to obtain spatially resolved spectroscopic information on the extended emission-line regions. In this paper we present the first results for SDSS J095100.54+051026.7, selected as a pilot target because it shows clearly separated, spatially extended ionised clouds visible already in the photometric data.

This paper is structured as follows. In Section~\ref{sec:obs} we describe the observations and the reduction procedure. Section~\ref{sec:analysis} presents the analysis of the long-slit and FPI data, including the emission-line fitting, kinematics, and diagnostic diagrams. Section~\ref{sec:conclusions} summarises the main results.

\section{Observation and data reduction}
\label{sec:obs}
As a pilot study, we focus here on the GBG candidate SDSS J095100.54+051026.7 (hereafter GP~117) at $z=0.24072$, selected from our sample of visually green galaxies
originally identified by participants of the Radio Galaxy Zoo project
\citep{Banfield2015}. To explore its properties, we obtained LS spectroscopy and scanning FPI, which together allow us to probe the spatial distribution and kinematics of the ionised gas. The details of the observational setup are summarised in Table~\ref{log}. The following subsections describe the observations and the data reduction procedures. The accepted spatial scale in this work is based on a flat $\Lambda$CDM cosmology with $H_0 = 70~\mathrm{km~s^{-1}~Mpc^{-1}}$, $\Omega_{\mathrm{m}} = 0.3$, and $\Omega_{\Lambda} = 0.7$. The conversion between angular and physical scales was performed using the \texttt{astropy.cosmology} package \citep{Astropy2013}. At the redshift of GP~117 ($z = 0.24072 \pm 0.00014$) as measured from the centroid positions of the \Ha, \Hb, \OIII, \NII, and \SII{} emission lines in our long-slit spectrum, this corresponds to a spatial scale of $3.8~\mathrm{kpc~arcsec^{-1}}$.

\begin{table}[htbp]
\caption{Log of GP~117 observations at the 6-m telescope.}
\label{log}
\begin{center}
\footnotesize
\renewcommand{\arraystretch}{1.25}
\begin{tabular}{lllllll}
\hline\hline
Data set & Date & $\text{T}_{exp}$, s & FOV & Seeing, $''$ & $\Delta \lambda$ \AA & FWHM, \AA \\
\hline
\multicolumn{7}{c}{LS spectroscopy} \\
SCORPIO-2 PA=$72^{\circ}$ & 2025 Jan 20 & 7200 & $1'' \times 6.8'$ & 1.7 & 3500--8500 & 7 \\
\multicolumn{7}{c}{FPI data} \\
SCORPIO-2 & 2025 Mar 07 & 3330  & $6.8' \times 6.8'$ & 2.3 & 6170--6300 & 13\\
\hline
\hline
\end{tabular}
\renewcommand{\arraystretch}{1}
\end{center}
\end{table}

\subsection{Long-slit spectroscopy}

\label{sec:LS}
We conducted long-slit spectroscopy using SCORPIO-2 multi-mode focal reducer \citep{Afanasiev2011} installed in the prime focus of the BTA 6-m telescope of Special Astrophysical Observatory of Russian Academy of Science (SAO RAS). 

Individual exposures were 600 s each, with a total exposure of 7200 s. The slit was positioned at PA=$72^{\circ}$, as shown in Fig.~\ref{fig:velocities} (top left). The observations were carried out on 2025 January 20. The log of observations with other parameters (spectral range -- $\Delta \lambda$, spectral resolution -- FWHM, field-of-view -- FOV) are given in Tab.~\ref{log}.

The primary reduction of the long-slit spectra was performed using the custom \texttt{longwid} pipeline (designed for SCORPIO-1,2) operating in the IDL environment. The reduction followed a standard sequence of steps to calibrate the data and extract 2D spectra.

The process began with the creation of master calibration frames: a master bias was constructed from multiple zero-exposure frames to model the bias readout level, and a master flat-field was produced by system of LEDs \citep{Afanasiev2017AstBu..72..458A}, providing stable and reproducible illumination for correcting variations of pixel-to-pixel sensitivity, optics transmission and non-uniformity of the slit width.

Wavelength calibration was performed by constructing a master He–Ne–Ar lamp frame from multiple comparison exposures and fitting a polynomial to positions of  identified lines, which was then applied to each science frame.
   
Science frames were subsequently corrected for the curvature of spectral lines across the detector, ensuring that each wavelength aligned along a constant spatial column. The geometric correction was derived from the He–Ne–Ar lamp calibration frames by tracing the positions of comparison lines along the slit and modelling their curvature to construct a distortion map, which was then applied to the science frames using a two-dimensional warping transformation. The frames were then divided by the normalized master flat to apply flat-fielding. The geometric distortions related with aberrations of SCORPIO-2 optics and misalignment between CCD chip horizontal axis and direction  of dispersion were modelled and corrected using a 13-point test pattern (13-dot mask) observed during the same run. This allowed for the precise rectification of the spectra.

The night sky background was modelled and subtracted by fitting a fifth-order polynomial to the spatial profile at each wavelength column in regions of the slit devoid of astronomical sources.  Following sky subtraction, residual cosmic ray hits were identified and removed with standard algorithms based on sigma-clipping algorithm applied to multiple exposures, as described for the CCD261-84 detector \citep[see Sec.~9 in][]{Afanasieva2023}. Residual cosmic rays were manually masked when necessary. 

Finally, the data were flux-calibrated. Spectrophotometric standard star Feige 56 was observed at a close zenith distance before the object to minimize the effects of atmospheric extinction variations. A sensitivity function was derived by comparing the extracted spectrum of the standard star to its absolute flux table. This function was applied to the science target to produce a final spectrum in absolute flux units [erg\,s$^{-1}$\,cm$^2$\AA$^{-1}$]. 

In parallel with the spectroscopic reduction, Poisson noise and readout noise were estimated from the raw frames. These measurements were then propagated to construct error frames, allowing proper error propagation throughout the subsequent data reduction steps.

From the long-slit data, we extracted one spectrum from the central region (hereafter \textit{C}) and two from the extended emission-line regions located to the East and West of the nucleus (\textit{E} and \textit{W} components, respectively). The extraction areas were chosen to match the regions of enhanced \OIII{} emission visible in the two-dimensional spectrum. The stellar population fitting was performed only for the central spectrum, while the emission-line analysis was carried out for all three components (Section~\ref{sec:fit}).

\subsection{3D spectroscopy with scanning FPI}

3D spectroscopic observations in the \OIII$\lambda5007$ emission line were performed with at the 6-m telescope with the SCORPIO-2 instrument in the scanning FPI mode having field of view $6.8 \times 6.8$ arcmin (Tab.~\ref{log}) with the scale $0.39''$per px. We used the low spectral resolution FPI20  that is usually employed as a tunable filter for mapping emission lines in various extended objects with MaNGaL photometer at the 1-m and 2.5-m telescopes \citep{Moiseev2020ExA....50..199M}. This FPI works in the order interference $\sim20$ (at 6560 \AA) and has a spectral resolution $\text{FWHM}\approx13$\AA\, ($\sim780\kms$ in the \OIII). Of course, this resolution is too low for detailed analysis of lines shape, however it is enough  to  study the gas kinematics if the amplitude of velocity changes exceeds $50\kms$, as it was shown in our previous research dedicated to Teacup AGN \citep{Moiseev2023Univ....9...66M}.

In contrast to `classical' observations with a high spectral resolution at which the full  range between adjacent interference orders was scanned \citep[][and references therein]{Moiseev2021AstBu..76..316M}, with low-resolution FPI20 we selected only 8 spectral channels for observations with the following wavelengths: 0, $\pm8$, $\pm16$, $\pm24$\AA{} relative to $6201$\AA{}. This central wavelength corresponds to the \OIII$\lambda5007$ emission line redshifted by the systemic velocity of the galactic nucleus, and $+100$\AA{} for the continuum emission. We sequentially switched corresponding gap between FPI plates (below -- `channel') with individual exposures 90 s per frame. In total we collect 3--6 frames in each channel, because only frames with the best seeing values were selected for the final analysis.
The medium band filter (central wavelength -- 6271 \AA, $\text{FWHM}=250$ \AA) isolated the  required spectral range with a single peak of FPI transmission \citep[see Fig.~1 in][]{Moiseev2020ExA....50..199M}. 

Internal  calibration included the exposures of the following sources: (i) continuous spectrum halogen lamp for creating a flat-field in the same channels as in the galaxy observations; (ii) the same He-Ne-Ar lamp as in the LS mode (Sec.~\ref{sec:LS}) scanned in the full working spectral range.  

The first steps of data reduction were  performed using  custom developed   IDL package \texttt{diwid}, specialized for reduction of direct images and MaNGaL data \citep{Moiseev2020ExA....50..199M}.

The reduction process began with the application of standard calibrations: subtraction of a master bias frame and division by a normalized master flat-field frame to correct for pixel-to-pixel sensitivity variations. Cosmic ray hits in individual frames   were removed using \texttt{L.A.COSMIC} program based on Laplacian edge detection  \citep{vanDokkum2001}. 

Subsequent steps involved the spatial alignment of all frames. This was achieved by matching the positions of foreground stars across individual exposures, and applying corresponding shifts to ensure precise registration throughout the data cube. Details obtained from stellar photometry were used to reject frames with poor seeing value, or with subsected excess of atmospheric extinction.

Individual exposures were then combined into data cube according to their channel numbers. Remaining cosmic ray hits were removed during the stacking process using a sigma-clipping technique.

Further data reduction steps were  based on algorithms from our   IDL package for FPI data reduction  \texttt{ifpwid} \citep[see references and description in ][]{Moiseev2021AstBu..76..316M}. The foreground air-glow emission were subtracted in the each data cube channel using a polynomial approximation of sky emission in a polar coordinate system in areas free from stars and contribution from galaxies. This procedure is important because the wavelength in the observed frames changes with radius \citep[see Fig.~4 in ][]{Moiseev2021AstBu..76..316M}. The correction on variations of atmospheric extinction from channel to channel were performed using photometry of stars with 2D Moffat function. Then the astrometric calibration was performed   via the astrometry.net\footnote{\url{http://astrometry.net/}} online service \citep{Lang2010AJ....139.1782L} and rotated to the custom orientation `North is up, East is to the left'.

Emission lines in the cube of  bias-subtracted frames illuminated by  He-Ne-Ar calibration lamps were fitted by Lorentz profile. Based on the fitting results  we created a  $\Lambda$-cube having the same size as the  object's data cube. The values in the  $\Lambda$-cube correspond   to wavelength of each pixel in the  object's data cube \citep{Moiseev2015AstBu..70..494M}. In other words, a non-linear  wavelength scale were created for each spaxel in field-of-view. Based on this wavelength calibration \OIII{} emission line was fitted with Voigt function that gives a good approximation of a FPI spectrum. The fitting results are images in the \OIII{} flux and line-of-sight velocity field shown in Fig.~\ref{fig:velocities} (bottom). Regions on the presented maps with signal-to-noise $S/N\le3$ were masked.
 
To flux-calibrate the data, the spectrophotometric standard BD+33d2642 was observed in the same instrumental configuration and atmospheric conditions. The identical reduction procedure was applied to the standard star observations. The calibration to the energetic units [erg\,s$^{-1}$\,cm$^2$] per pixel or arcsec$^2$ was performed similar with MaNGaL data as desribed in \citep{Moiseev2020ExA....50..199M}. The instrumental response curve was derived by comparing the observed and tabulated fluxes of the standard star, after correcting for atmospheric extinction, and was subsequently applied to the object.

\section{Data analysis}
\label{sec:analysis}
The combined use of long-slit spectroscopy and FPI mapping provides a methodology for the spatially resolved analysis of the ionized gas in GP~117.

\subsection{Stellar population subtraction and emission-line fitting}
\label{sec:fit}

A crucial step in the analysis of spectra of galaxies with weak or fading AGN signatures is the separation of stellar and gaseous contributions in the central regions, where the host galaxy continuum can significantly dominate. To account for the stellar population, we employed the \texttt{pPXF} (penalized pixel-fitting) package \citep{Cappellari2017}, fitting the observed spectra with the eMILES stellar population synthesis (SPS) models \citep{Vazdekis2016}, without including an AGN component. Although the nuclear contribution is relatively weak, consistent with a fading or low-luminosity AGN scenario, the emission lines in the nuclear spectrum are still noticeable. Therefore, the stellar continuum fitting was performed iteratively: the strongest emission-line regions were first masked, followed by the fitting of the absorption features and subsequent refinement of the emission-line model until convergence was reached. This approach ensures a reliable subtraction of the absorption features, particularly those affecting the Balmer lines. The resulting luminosity-weighted stellar age and metallicity for this object are $\langle Age \rangle = 12.6 \pm 0.5\, Gyr$ and $\langle \mathrm{[M/H]} \rangle = -0.40 \pm0.05$, respectively.

After continuum subtraction, the residual emission-line spectra were fitted with Gaussian profiles using the standard \texttt{python} library \texttt{scipy}. From this fitting, we derived the radial velocities and line fluxes for each emission line, which are essential for constructing diagnostic diagrams that distinguish between different ionisation mechanisms and trace the fading signatures of AGN activity (see Fig.~\ref{fig:diagnostic}).

\subsection{Spatially Resolved Kinematics} 
\begin{figure}[thp]
\centerline{\includegraphics[width=0.94\textwidth,clip=]{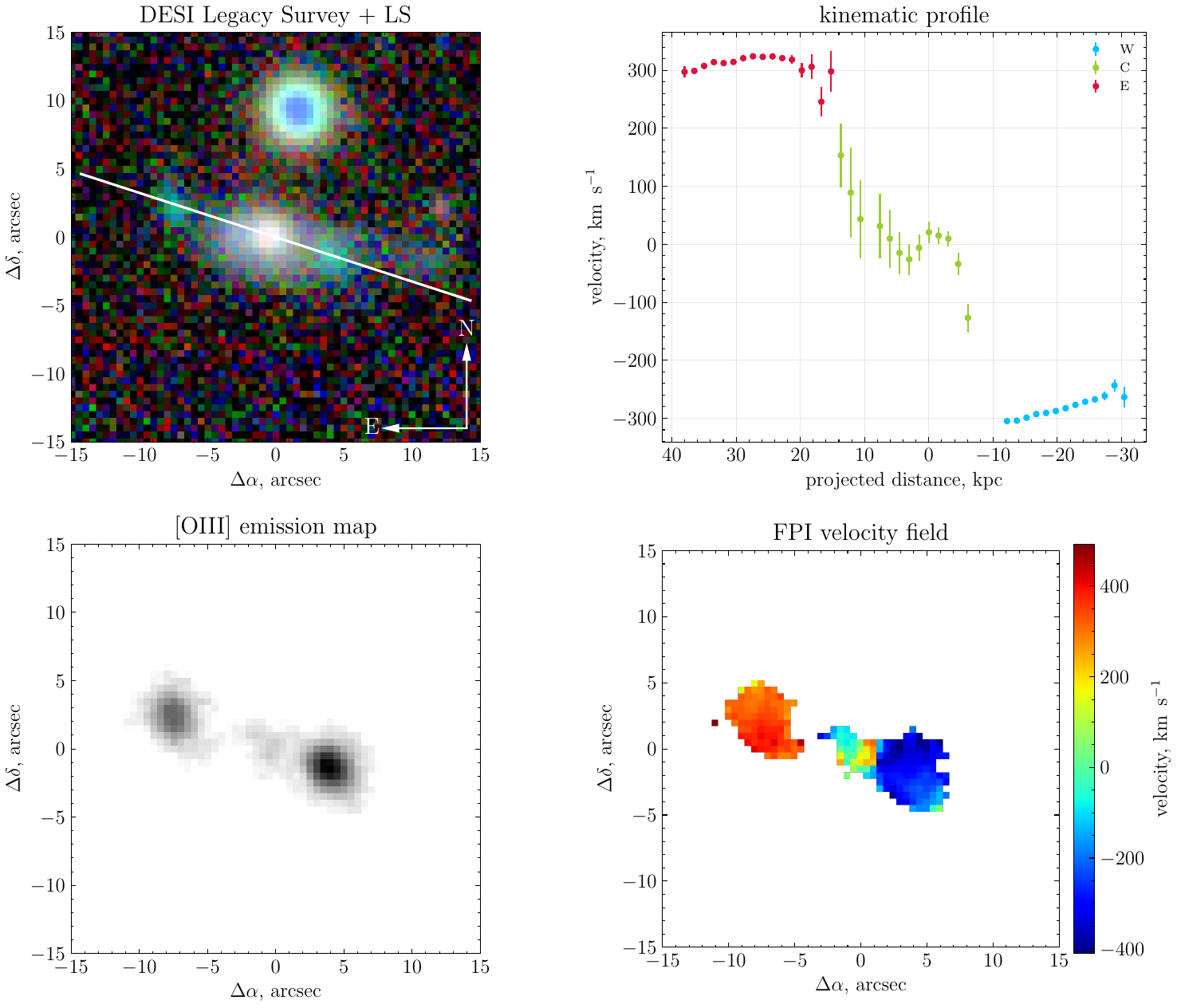}}
\caption{\textit{(Top left:)} Optical image from the DESI Legacy Survey. The white line indicates the position of the LS at PA = 72°, used for the spectroscopic observations. \\
\textit{(Top right:)} Line-of-sight velocity distribution derived from the long-slit spectroscopic data. Velocity distribution corrected for the systemic velocity $v_{sys}=71567\, km\,s^{-1}$. The green symbols represent the stellar-population-subtracted gas component in the central galactic region, while the red and blue symbols trace the velocity and its measurement errors in the extended clouds on either side of the galaxy. \\
\textit{(Bottom:)} Maps obtained from Voigt fitting of FPI spectra: flux map in the \OIII$\lambda$5007 emission line \textit{(left)} and corresponding line-of-sight velocity field \textit{(right)}. \label{fig:velocities}
}
\end{figure}

The long-slit spectra provide the primary data for kinematic analysis. Velocities were derived by fitting Gaussian profiles to the main emission lines after subtraction of the stellar continuum. The line-of-sight velocity distribution shown in Fig.~\ref{fig:velocities} (top right). Furthermore, the \OIII{} velocity field was reconstructed from the FPI data cube by fitting the \OIII{} line profile in each spatial element with a Voigt function, producing the two-dimensional map of line-of-sight velocities (Fig.~\ref{fig:velocities}) (bottom right). A more comprehensive interpretation of the kinematics will be addressed in a forthcoming study, where we will combine these data with a larger sample and additional multiwavelength observations.

\subsection{Excitation and Ionization Conditions} 
The primary tool for diagnosing the mechanisms powering the ionized gas (e.g., star formation vs. active galactic nuclei) is the analysis of emission-line ratios.

\begin{figure}[thp]
\centerline{\includegraphics[width=\textwidth,clip=]{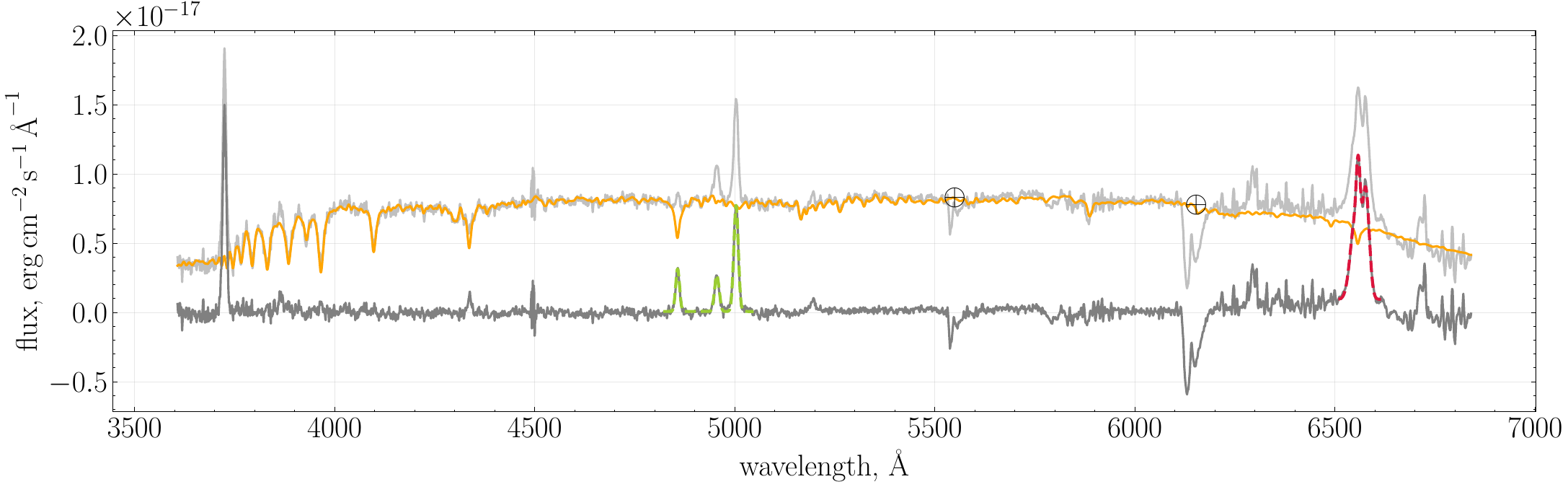}}
\vspace{0.5cm}
\centerline{\includegraphics[width=5.cm,clip=]{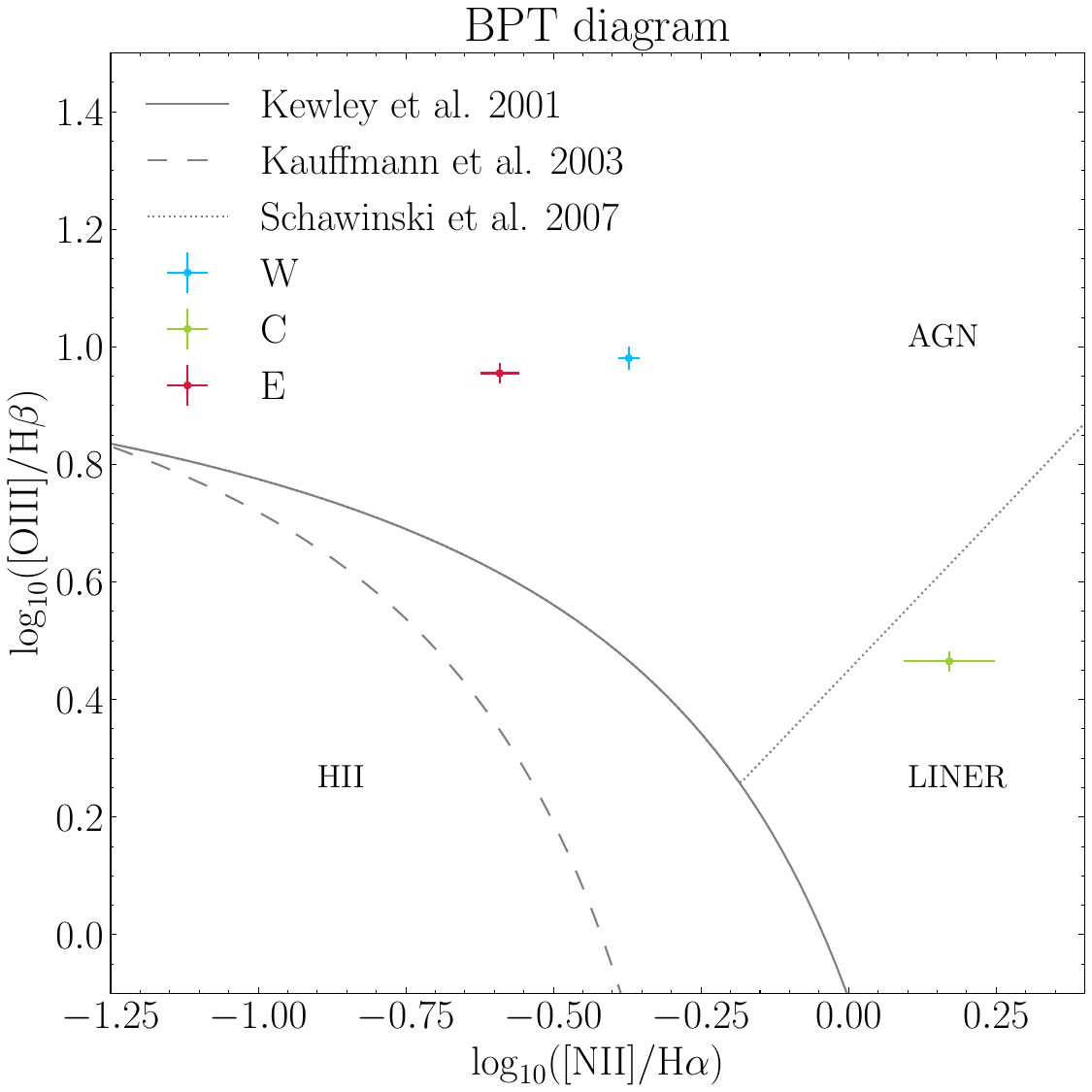} 
           \hspace{0.5cm}
           \includegraphics[width=5.cm,clip=]{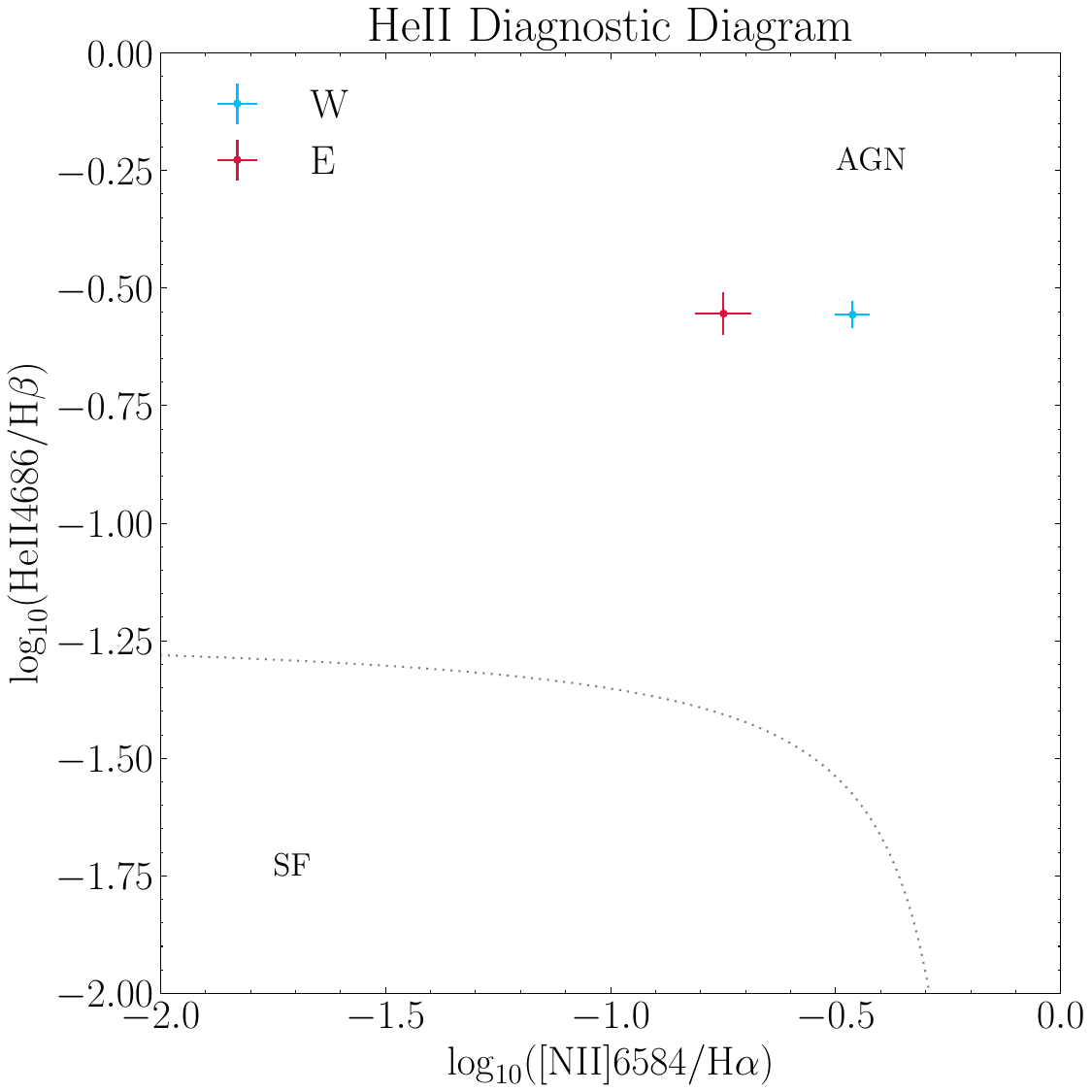}
          }
\caption{
\textit{(Top):} Nuclear spectrum of GP~117. The observed SCORPIO-2 spectrum is shown in light gray, the stellar population model — in orange. The dark gray line corresponds to the residual spectrum after subtracting the stellar continuum. The green dashed and red dashed lines represent Gaussian fits to the \OIII+\Hb\, and \NII+\Ha\, emission lines, respectively. Telluric absorption lines are marked with the $\oplus$ symbol at the corresponding wavelengths. \\
\textit{(Bottom left):} Standard \OIII$\lambda$5007/\Hb\, vs. \NII$\lambda$6583/\Ha\, (BPT) diagram. The solid and dashed curves represent the separation lines between different ionization sources \citep{Kewley2001, Kauffmann2003, Schawinski2007}. \\
\textit{(Bottom right):} \HeII/\Hb\, vs. \NII$\lambda$6583/\Ha\, diagram \citep{Shirazi2012}. The green point marks the central galactic region, while the red and blue points show the integrated line ratios for the extended clouds.}
\label{fig:diagnostic}
\end{figure}

\textit{BPT Diagnostics:} From the long-slit spectra, we measure the fluxes of key diagnostic lines (\OIII$\lambda$5007, \Hb, \Ha, \NII$\lambda$6583) at various positions. These ratios are plotted on standard Baldwin-Phillips-Terlevich (BPT) diagrams \citep{Baldwin1981} in Fig.~\ref{fig:diagnostic}. The central region (C) of the galaxy lies within the LINER area of the diagram. The observed emission-line widths ($\sim$330–400~km~s$^{-1}$) show no evidence for the extremely broad profiles expected from strong shock ionization. While some local kinematic disturbances may be present, photoionization by the AGN appears to be the dominant excitation mechanism. Meanwhile, the extended ionised clouds (W and E) exhibit systematically higher ionisation, which is indicative of a previous phase of AGN activity.

\textit{Helium Line Analysis:} 
For the \HeII, emission line in the long-slit spectra, we construct a helium diagnostic diagram (\HeII/\Hb , vs. \NII$\lambda$6583/\Ha) \citep{Shirazi2012} to probe the ionising source and the physical conditions of the gas (Fig.~\ref{fig:diagnostic}). Notably, \HeII\ emission is detected predominantly in the extended ionised clouds, while it is absent in the central region, highlighting that the hard ionising radiation from the AGN primarily affects the outer regions.

These results confirm the presence of large-scale ionised clouds surrounding GP~117, which were previously only suspected based on photometric data. The \OIII$\lambda$5007 emission extends up to $\sim 40$ kpc from the nucleus, as seen from both the line-of-sight velocity curve and the [O III] flux map (Fig.~\ref{fig:velocities}) (right panels).

\section{Conclusions}
\label{sec:conclusions}

In this work, we present a pilot study of the Green Bean galaxy GP~117 (SDSS J095100.54+051026.7), based on new long-slit and 3D spectroscopic observations obtained with the SCORPIO-2 spectrograph at the 6-m Russian telescope. These data provide the first spectroscopic confirmation of the large-scale ionised clouds in this system, allowing us to characterise their ionisation state and kinematics in detail. The \OIII$\lambda$5007 emission is traced up to $\sim 40$~kpc from the nucleus on each side, confirming the presence of extended high-ionisation gas previously suspected only from imaging data. 

The properties of these ionised clouds show similarities to other systems with extended high-ionisation emission, such as Hanny’s Voorwerp in the local Universe~\citep{Lintott2009} and several Green Bean galaxies analysed previously~\citep{Schirmer2016}. In these objects, bright \OIII{} emission on tens-of-kiloparsec scales and enhanced ionisation away from the nucleus have been reported as common features. Although the physical interpretation of such structures varies across individual cases, GP~117 exhibits ionisation characteristics and spatial morphology comparable to those seen in these previously studied systems.

To quantify these properties, we modelled the stellar population using the \texttt{pPXF} package with eMILES templates to subtract the host galaxy continuum in the central regions, allowing reliable recovery of emission-line fluxes. Gaussian fitting of the emission lines then provided kinematic parameters and fluxes for key diagnostic features, enabling the construction of classical and extended diagnostic diagrams. Resolved emission-line ratio maps and spatially resolved kinematics suggest a radiative fading scenario in GP~117, and our next step is to combine this optical analysis with radio observations to gain a deeper understanding of the evolutionary status of GP~117 and related systems.


\acknowledgements
This study was supported by the Russian Science Foundation, project no. 25-12-00129 ‘Study of activity cycles of galactic nuclei’. The observational data were  obtained with the 6-m telescope of the Special Astrophysical Observatory of the Russian Academy of Sciences carried out with the financial support of the Ministry of Science and Higher  Education of the Russian Federation. The renovation of the telescope equipment is currently provided within the national project ``Science and Universities''.

\bibliography{Arshinova_caosp310}

\end{document}